%% file: main.tex
\documentclass[conference]{IEEEtran}
\IEEEoverridecommandlockouts

\usepackage[utf8]{inputenc} 
\usepackage{cite}           
\usepackage{amsmath,amssymb,amsfonts} 
\usepackage{graphicx}       
\usepackage{textcomp}       

\usepackage[table,xcdraw,dvipsnames]{xcolor} 
\usepackage{multirow}       
\usepackage{array}          
\usepackage{tabularx}       
\usepackage{booktabs}       
\usepackage{colortbl}       

\usepackage{algorithm}      
\usepackage{algpseudocode}  

\usepackage{subcaption}     
\usepackage{subfig}         
\usepackage{dblfloatfix}    

\usepackage{tikz-qtree}     
\usetikzlibrary{shapes,arrows} 

\usepackage{cleveref}       
\usepackage{adjustbox}      
\usepackage{commath}        
\usepackage{pbalance}       
\usepackage{comment}        
\usepackage{listings}       
\usepackage{csquotes}       
\usepackage{gensymb}        

\usepackage[labelformat=simple]{caption}

\algnewcommand\algorithmicswitch{\textbf{switch}}
\algnewcommand\algorithmiccase{\textbf{case}}
\algnewcommand\algorithmicassert{\texttt{assert}}
\algnewcommand\Assert[1]{\State \algorithmicassert(#1)}

\def\BibTeX{{\rm B\kern-.05em{\sc i\kern-.025em b}\kern-.08em
    T\kern-.1667em\lower.7ex\hbox{E}\kern-.125emX}}

\title{

MATTER: \underline{M}ULTI-STAGE \underline{A}DAPTIVE \underline{T}HERMAL \underline{T}ROJAN
FOR \underline{E}FFICIENCY \& \underline{R}ESILIENCE DEGRADATION \vspace{-2mm}

\thanks{The authors would like to thank the following funding agencies; XYZ Grants \# XXXXXX and \# XXXXXX.}
}

\begin{document}


 \author{\IEEEauthorblockN{Mehdi Elahi$^1$, Mohamed R. Elshamy$^2$, Abdel-Hameed Badawy$^2$, Mahdi Fazeli$^3$, Ahmad Patooghy$^1$}
 $^1$ Department of Computer Systems Technology, North Carolina A\&T State University, NC, 27411\\
$^2$ New Mexico State University, United states\vspace{-4mm}
$^3$ Halmstad University, Sweden\vspace{-2mm}
 }

\maketitle

\sloppy
\begin{abstract} 

As mobile systems become more advanced, the security of System-on-Chips (SoCs) is increasingly threatened by thermal attacks. This research introduces a new attack method called the Multi-stage Adaptive Thermal Trojan for Efficiency and Resilience Degradation (MATTER). MATTER takes advantage of weaknesses in Dynamic Thermal Management (DTM) systems by manipulating temperature sensor interfaces, which leads to incorrect thermal sensing and disrupts the SoC's ability to manage heat effectively. Our experiments show that this attack can degrade DTM performance by as much as 73\%, highlighting serious vulnerabilities in modern mobile devices. By exploiting the trust placed in temperature sensors, MATTER causes DTM systems to make poor decisions i.e., failing to activate cooling when needed. This not only affects how well the system works but also threatens the lifespan of the hardware. This paper provides a thorough analysis of how MATTER works and emphasizes the need for stronger thermal management systems in SoCs.

\end{abstract}

\begin{IEEEkeywords}
Thermal trojan Attacks, Dynamic Thermal Management (DTM), System-on-Chips (SoCs)
\end{IEEEkeywords}

\input{Sources/1-Intro.tex}

\input{Sources/2-Background.tex}

\input{Sources/3-MATTER.tex}

\input{Sources/4-EVALUATION.tex}

\input{Sources/5-ConC.tex}




\end{document}

%% file: Sources/1-Intro.tex
\vspace{-1mm}
\section{Introduction}
\label{sec:Intro}


System-on-Chips (SoCs) have been widely used as the backbone of high-performance mobile systems in myriad of tasks integral to the human's daily life \cite{jeong2022band,10356076,patooghy2019your}. However, 
mobile SoCs face significant thermal-power constraints budget \cite{wang2023efficient} primarily due to their compact form factors and limited cooling capabilities in handheld devices like smartphones \cite{7100138, 8509120 }. Ensuring user comfort by mitigating skin temperature to prevent discomfort or burning sensations adds another layer of complexity to the thermal management puzzle \cite{gong2023performance,8671643}. 
Hence, Dynamic Thermal Management (DTM) techniques have been widely utilized among mobile systems to effectively regulate and minimize high operating temperatures 

The accuracy and reliability of thermal sensors (that DTM methods rely on) are crucial for effective thermal management strategies \cite{sonmez2017compact, ding2020novel}. The sensor malfunctioning can occur due to a variety of factors, ranging from unintentional faults to deliberate tampering. Unintentional faults may include issues arising from fabrication defects, aging effects, susceptibility to noise, and process variations\cite{Oukaira2022FEMbasedTP,jlpea4040304}. These factors can lead to inaccuracies in thermal measurements, which can in turn, impact the overall performance and reliability of the system. 
Additionally, there are significant security risks associated with deliberate tampering, such as the insertion of Hardware Trojans (HT), which are malicious modifications trying to alter the behavior of thermal sensors, leading to falsified temperature values \cite{hasegawa2023node}. Erroneous temperature readings can trigger unnecessary frequency reduction or core throttling by DTM, with the aim of reducing the chip's performance \cite{li2020accurate}. Furthermore, incorrect thermal data can accelerate the chip's aging process, thereby reducing its lifespan. The insertion of thermal HTs thus poses a significant threat to the security and reliability of thermal management systems \cite{cruz2022automatic}.

There are a few research papers in the literature that tried to address malfunctioning thermal sensors. Authors in \cite{abdelrehim2022bic} introduced the Blind Identification Countermeasure (BIC), a technique derived from the Blind Power Identification (BPI) algorithm \cite{said2018understanding}. BIC aims to detect, contain, and isolate malicious sensors, thereby offering accurate temperature estimations and safeguarding chip integrity. However, the effectiveness of such methods is constrained, as they are vulnerable to more advanced and recently developed attacks. In this paper, we propose a novel attack that does not involve manipulating thermal data which is easy to detect, but instead alters the temperature sensed by the DTM system, thereby disrupting its operation.  
The contributions of this paper are as follows.
\begin{itemize}
    \item We introduce MATTER, a novel thermal Trojan attack that exploits vulnerabilities in the DTM thresholds. By subtly influencing the sensed thermal profile, the attack can bypass the system's defenses, causing it to mismanage thermal events, potentially leading to system instability, performance degradation, or even hardware damage without detection. 
    
    \item Instead of attacking thermal sensors, we target the DTM sensed temperature interface. To the best of our knowledge, this is the first attempt to target the sensed data within the DTM system, rather than focusing on the temperature sensors themselves.
    
    \item We run a set of heterogeneous and homogeneous configurations to comprehensively evaluate the MATTER behavior for compute-intensive, memory-intensive, and mixed workloads. 
\end{itemize}

The paper is organized as follows: \Cref{sec:Background} provides background and prior research, \Cref{Sec:Proposed} discusses the proposed Attack (MATTER), Simulation results are gathered in \Cref{Sec:EvALUATION}, and finally \Cref{Sec:ConC} concludes the paper.

%% file: Sources/2-Background.tex
\section{BACKGROUND AND RELATED WORKS}
\label{sec:Background}

Thermal management is a critical aspect of modern SoC design, particularly for mobile devices. DTM systems employ multiple temperature thresholds to maintain safe operating conditions while optimizing performance \cite{coskun2008proactive}.
The \textit{trigger threshold}, $Th_{Trigger}$, serves as an early warning that  initiates corrective actions like adjusting operational frequency (through Dynamic Voltage and Frequency Scaling, DVFS) when temperature approaches unsafe levels. This helps prevent overheating while minimizing performance impact. The \textit{critical threshold}, $Th_{Critical}$, representing the highest thermal limit, triggers aggressive interventions such as significant voltage/frequency reductions, core throttling, or emergency shutdowns to prevent hardware damage. Once temperatures fall below the \textit{recovery threshold}, $Th_{Recovery}$, the system gradually resumes normal operations, carefully balancing performance and thermal management.

The manipulation of thermal sensors in mobile SoCs has emerged as a significant security concern. Researchers have explored various attack vectors and countermeasures to address this issue. Sharifi and Rosing \cite{sharifi2010accurate} proposed an indirect temperature sensing approach using a Kalman filter. Their study evaluated the performance improvements achieved by avoiding false throttling events triggered by inaccurate temperature data. Building on this work, Yan et al. \cite{yan2021chip} presented an on-chip HT detection scheme utilizing temperature sensors and thermal profiles. They explored two HT-agnostic Machine Learning (ML) classifiers, achieving detection rates of 55\% and 95\% for Minimum Volume Enclosing Ellipsoid (MVEE) and One-Class Support Vector Machine (1-SVM), respectively. 

Guo et al. \cite{guo2020securing} introduced a detection technique that leverages the time it takes for ICs to reach steady-state temperatures to identify HTs. Their method demonstrated detection rates ranging from 84\% to 100\%, with false positive rates between 1\% and 18\%. These approaches highlight the growing interest in thermal-based security measures for SoCs. More recently, the Blind Identification Countermeasure (BIC) \cite{abdelrehim2022bic} has been proposed as a comprehensive approach to detect and mitigate malicious thermal sensor attacks in mobile SoCs. During four phases, the BIC establishes a thermal resistance transfer matrix during offline testing, continuously monitors the chip during runtime, identifies malicious sensors, and estimates accurate core temperatures. While BIC presents an interesting method for safeguarding mobile SoCs against sophisticated thermal attacks, the attack model used to test it has been criticized for its simplicity. The HT used in the test could be detected by simply comparing average CPU temperatures with and without the HT present, and the performance degradation caused by CPU throttling was not quantitatively evaluated. 

While existing thermal-based security measures demonstrate promising methods to identify and mitigate thermal sensor attacks in mobile SoCs, significant challenges remain. Most approaches primarily focus on detecting hardware Trojans (HTs) or irregular temperature profiles, but they do not fully address vulnerabilities within the DTM thresholds. Specifically, current methods do not explore how these thresholds could be directly exploited to disrupt DTM's adaptive mechanisms without manipulating thermal sensors. In light of this gap, our research introduces a novel attack targeting DTM thresholds and demonstrates how this approach exploits its policies.

%% file: Sources/3-MATTER.tex
\section{The Proposed Attack (MATTER)}
\label{Sec:Proposed}

In this section, we exploit the DTM vulnerability to construct the Multi-stage Adaptive Thermal Trojan for Efficiency \& Resilience degradation (MATTER). First, we present the threat model and assumptions. Then, we introduce MATEER and last the "thread biasing" technique as the heating-up process for triggering MATTER is discussed.

\subsection{Threat Model and Assumptions}
\label{Sec:Threat Model}

HTs represent a significant security threat to integrated circuits, with multiple potential insertion vectors throughout the design and manufacturing process. These vulnerabilities include insider threats from untrusted team members who could directly manipulate circuit designs \cite{sebt2018circuit}, the incorporation of potentially compromised third-party IP (3PIP) cores that may not have undergone thorough security audits \cite{zhang2014detrust}, and the use of untrusted design tools \cite{mukhopadhyay2014hardware} that could be exploited to inject malicious functionality. In the specific context of sensor-based systems, HTs can be particularly insidious. For example, a third-party processing core with an integrated thermal sensor presents an opportunity for malicious insertion. The core designers, armed with detailed knowledge of heat-generating components, could strategically place compromised thermal sensors near these hotspots to maximize their effectiveness \cite{abdelrehim2022bic}. In our threat model, we assume that the X86-based system configuration remains intact while the attacker gains access to local resources, i.e., system interfaces.

\begin{figure}[t]
    \centering
     \vspace{-3mm}
    \includegraphics[scale=0.53]{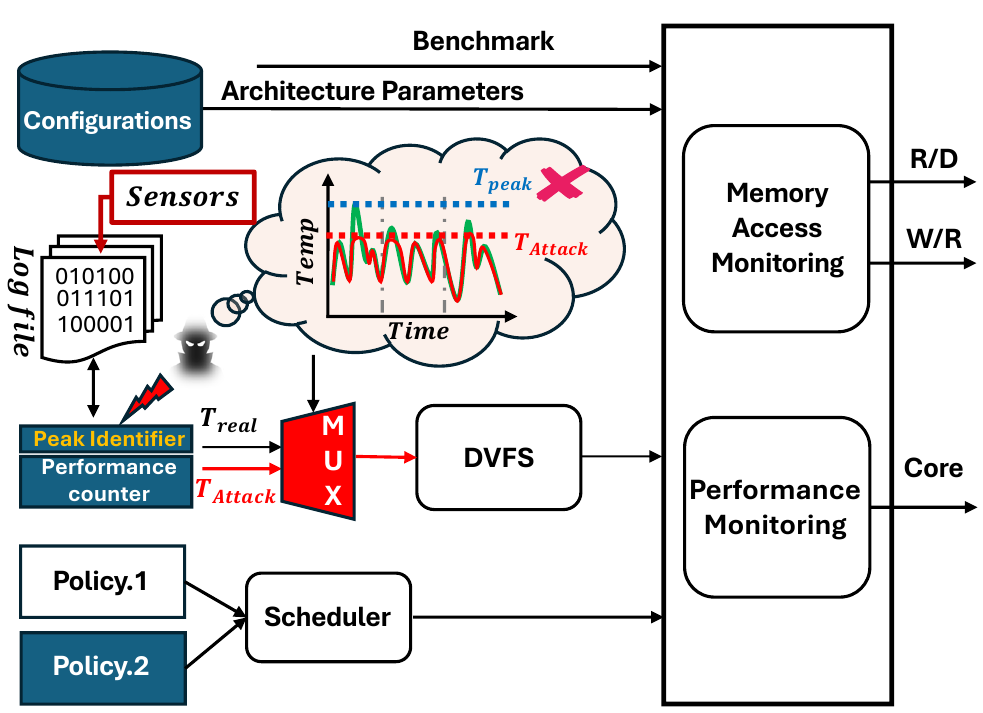} 
    \caption{Illustration of Attack on the System's Interface} 
    \label{fig:AttackDiagram}
\end{figure}

\begin{table*}[b]
\caption{Real System response With and without the attack}
\label{tab:AttackSample}
\centering
\begin{tabular}{cccccc}
\rowcolor[HTML]{C0C0C0} 
\multicolumn{6}{c}{\cellcolor[HTML]{C0C0C0}{\textbf{Stage.1: Real System Reaction With and without the attack}}} \\ \hline
\rowcolor[HTML]{EFEFEF} 
\multicolumn{1}{|c|}{\cellcolor[HTML]{EFEFEF}Time} &
  \multicolumn{1}{c|}{\cellcolor[HTML]{EFEFEF}Real Temp. (C)} &
  \multicolumn{1}{c|}{\cellcolor[HTML]{EFEFEF}Attacked Temp (C)} &
  \multicolumn{1}{c|}{\cellcolor[HTML]{EFEFEF}Frequency without DVFS} &
  \multicolumn{1}{c|}{\cellcolor[HTML]{EFEFEF}DVFS's reaction without Attack} &
  \multicolumn{1}{c|}{\cellcolor[HTML]{EFEFEF}DVFS's reaction With Attack} \\ \hline
\rowcolor[HTML]{FFFFFF} 
\multicolumn{1}{|c|}{\cellcolor[HTML]{FFFFFF}0} &
  \multicolumn{1}{c|}{\cellcolor[HTML]{FFFFFF}77.5} &
  \multicolumn{1}{c|}{\cellcolor[HTML]{FFFFFF}78} &
  \multicolumn{1}{c|}{\cellcolor[HTML]{FFFFFF}4 GHz} &
  \multicolumn{1}{c|}{\cellcolor[HTML]{FFFFFF}3.2 GHz} &
  \multicolumn{1}{c|}{\cellcolor[HTML]{FFFFFF}3.4 GHz} \\ \hline
\multicolumn{1}{|c|}{\cellcolor[HTML]{EFEFEF}10} &
  \multicolumn{1}{c|}{\cellcolor[HTML]{EFEFEF}77.8} &
  \multicolumn{1}{c|}{\cellcolor[HTML]{EFEFEF}77.8} &
  \multicolumn{1}{c|}{\cellcolor[HTML]{EFEFEF}4 GHz} &
  \multicolumn{1}{c|}{3.4 GHz} &
  \multicolumn{1}{c|}{\cellcolor[HTML]{EFEFEF}3.4 GHz} \\ \hline
\rowcolor[HTML]{FFFFFF} 
\multicolumn{1}{|c|}{\cellcolor[HTML]{FFFFFF}20} &
  \multicolumn{1}{c|}{\cellcolor[HTML]{FFFFFF}77.9} &
  \multicolumn{1}{c|}{\cellcolor[HTML]{FFFFFF}80.4} &
  \multicolumn{1}{c|}{\cellcolor[HTML]{FFFFFF}4 GHz} &
  \multicolumn{1}{c|}{\cellcolor[HTML]{FFFFFF}3.6 GHz} &
  \multicolumn{1}{c|}{\cellcolor[HTML]{FFFFFF}3.8 GHz} \\ \hline
\rowcolor[HTML]{C0C0C0} 
\multicolumn{6}{|c|}{\cellcolor[HTML]{C0C0C0}{\textbf{Stage.2: Real System Reaction With and without the attack}}} \\ \hline
\multicolumn{1}{|c|}{30} &
  \multicolumn{1}{c|}{80.5} &
  \multicolumn{1}{c|}{79.4} &
  \multicolumn{1}{c|}{4 GHz} &
  \multicolumn{1}{c|}{1 GHz} &
  \multicolumn{1}{c|}{4 GHz} \\ \hline
\rowcolor[HTML]{EFEFEF} 
\multicolumn{1}{|c|}{\cellcolor[HTML]{EFEFEF}40} &
  \multicolumn{1}{c|}{\cellcolor[HTML]{EFEFEF}81.9} &
  \multicolumn{1}{c|}{\cellcolor[HTML]{EFEFEF}79.8} &
  \multicolumn{1}{c|}{\cellcolor[HTML]{EFEFEF}4 GHz} &
  \multicolumn{1}{c|}{\cellcolor[HTML]{EFEFEF}1 GHz} &
  \multicolumn{1}{c|}{\cellcolor[HTML]{EFEFEF}4 GHz} \\ \hline
\multicolumn{1}{|c|}{50} &
  \multicolumn{1}{c|}{82.6} &
  \multicolumn{1}{c|}{79.6} &
  \multicolumn{1}{c|}{4 GHz} &
  \multicolumn{1}{c|}{1 GHz} &
  \multicolumn{1}{c|}{4 GHz} \\ \hline
\end{tabular}
\end{table*}

\Cref{fig:AttackDiagram} illustrates the interface targeted by MATTER, demonstrating the adversarial access and intentions imposed on this critical component. Specifically, a performance counter periodically reads temperature data from thermal sensors and records it in a log file, capturing real-time thermal conditions across system components. A peak identifier module then processes this data to identify the maximum temperature value among the components. This peak temperature is subsequently used to regulate the DVFS mechanism, which adjusts the system's voltage and frequency levels based on current thermal conditions to optimize performance while preventing overheating. In this work, we focus on the interface linking the peak identifier to the DVFS control module, as it is a crucial yet vulnerable point of interaction. By subtly manipulating the peak identifier architecture, it returns the attacker's desired temperature instead of the real temperature's peak. This approach is taken to keep temperature reading untouched, outperforming temperature-based countermeasures against such thermal attacks. 


\subsection{Attack Scenario} 
\begin{figure}[t]
    \centering
    \includegraphics[scale=0.42]{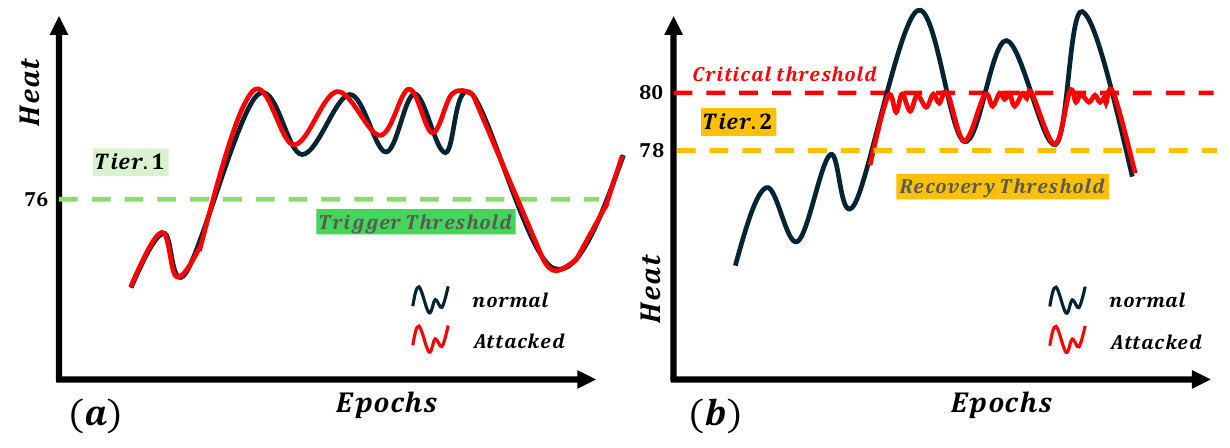} 
    \caption{MATTER's behavior across two defined stages while DTM is deactivated, activated, and under attack} 
    \label{fig:AttackScenario}
\end{figure}

In this section, we present a novel Trojan attack termed the Multi-stage Adaptive Thermal Trojan for Efficiency and Resilience Degradation (MATTER). As outlined previously in \cref{Sec:Threat Model}, this attack targets the temperature-sensing interface of the DTM system to provoke inaccurate responses based on DTM's critical threshold levels. MATTER strategically targets these thresholds to degrade system efficiency and resilience, ultimately compromising thermal stability and control. This purpose is served by manipulating the DTM interface's temperature sensing in two distinct stages:

\textbf{Stage-1}: In the first stage, called the \textbf{"trigger-crossing interval"}, the attack subtly raises the temperature by a small amount-between 0.5\degree C and 1\degree C (\Cref{fig:AttackScenario}.a). This minor increase is designed to look like a natural temperature drift, making the system treat it as a harmless change. Because of this, DVFS responds by optimizing performance, leading to a slight reduction in both voltage and frequency compared to when DVFS is not active. However, as shown in \cref{tab:AttackSample}, without the attack, DTM would typically run the system at a lower frequency during the same period. The attack carefully calibrates this temperature adjustment to avoid detection, allowing the system to appear as if it is operating normally, while actually maintaining performance at a higher level than it would operate otherwise.


\textbf{Impacts of Stage-1:} 
This initial manipulation stage carefully conditions the system by subtly adjusting DTM's thermal sensing to create an artificial baseline, setting the groundwork for sustained, unnoticed stress on the device. By gradually modifying the expected thermal behavior, the system enters a state of deceptive stability that does not trigger any immediate alerts, fostering a false sense of operational normalcy. Additionally, During this phase, the attack is strategically accumulating "temperature credits," a form of deferred thermal budget that allows the system to operate under slightly cooler conditions than normal, when it is needed. These credits are essential for the next stage, as they effectively "store" thermal allowance that we can exploit later. Specifically, to counter methods that monitor average temperature and detect attacks, we ensure that the temperature appears normal during this phase, making it harder for these detection mechanisms to identify anomalies. In the subsequent attack phase, or Attack Stage 2, these accumulated credits are "spent" to prolong high-operating conditions while keeping the temperature lower than its actual value and below the critical threshold. This approach enables the attack to sustain elevated stress levels on the system by avoiding the system's usual thermal management responses.

\textbf{Stage-2: }In the second stage, \textbf{"critical-crossing interval"}, the attack becomes more sophisticated. The DTM's sensed temperature fluctuates within a narrow range, typically between -1.5\degree C and -2\degree C, following a normal distribution and simulating natural sensor variations as follows: 

\begin{equation}
\begin{split}
    \text{Attacked} &= \text{Current\_Temp} - \mathcal{N}([1.5, 2])   \\
    \text{Attacked\_Temp}[i] &= \min \left( \text{Attacked}, \text{Critical} - \mathcal{N}([0.1, 0.3]) \right)
\end{split}
\end{equation}
This stage strategically maintains the DTM's sensed temperature just below the critical threshold (e.g., 80\degree C - see \Cref{fig:AttackScenario}.b), preventing the DVFS from engaging its protective measures, i.e., core throttling.

\textbf{Impacts of Stage-2: }
As illustrated in the Attack stage-2 section of \Cref{tab:AttackSample}, this manipulation forces the system to continue operating at maximum frequency (e.g., 4 GHz) while the real temperature has already surpassed the critical threshold. In the absence of this attack, DVFS would typically force the frequency down to around 1 GHz in response to the high temperature, effectively throttling the system to prevent overheating. By keeping the DTM's sensed temperature below this range, this attack bypasses these preventative adjustments, inducing prolonged operation at higher, potentially unsafe frequency (i.e., 4GHz) thus resulting in higher power consumption and temperature levels.

\subsection{MATTER's Triggering Mechanism}
\label{Sec:thread biasing}

The MATTER attack employs a triggering technique called "Thread Biasing" to extract critical side-channel information about the target system's thermal management parameters, specifically the thresholds $Th_{Critical}$, $Th_{Trigger}$, and $Th_{Recovery}$. To execute Thread Biasing, the attacker leverages the "/system/thread\_manager" interface to allocate tasks, directing an increased workload to a specific core, i.e., $i$, in the attack model. This targeted approach elevates the core's temperature, forcing the system to activate its DTM features. The controlled heating process enables the attacker to monitor the system's responses, including thermal throttling and recovery to normal operating temperatures. By studying the system's thermal behavior under these conditions, the attacker gains insights into the system's thermal management strategies and thresholds. 

For example, the attacker might observe that when the temperature of core $"i"$ reaches 85\degree C, there is a drastic reduction in its operating frequency from 3.5 GHz to 2.8 GHz. This sudden drop indicates the throttling threshold ($Th_{Critical}$). As the workload is maintained, the attacker notices that the frequency remains at 2.8 GHz until the temperature drops to 75\degree C, at which point the frequency returns to 3.5 GHz. This temperature marks the recovery threshold ($Th_{Recovery}$). By repeating this process multiple times and analyzing the data, the attacker can accurately determine: The throttling threshold: 85\degree C, The recovery threshold: 75\degree C, and The extent of frequency reduction: 20\% (from 3.5 GHz to 2.8 GHz). This information provides valuable insights into the system's thermal management strategy, allowing the attacker to fine-tune their approach for MATTER's Triggering Mechanism.

%% file: Sources/4-EVALUATION.tex
\section{EvALUATION}
\label{Sec:EvALUATION}

\subsection{Performance impact}
\label{Performance impact}

\begin{figure*}[t]
    \centering
    \centering
    \includegraphics[scale=0.51]{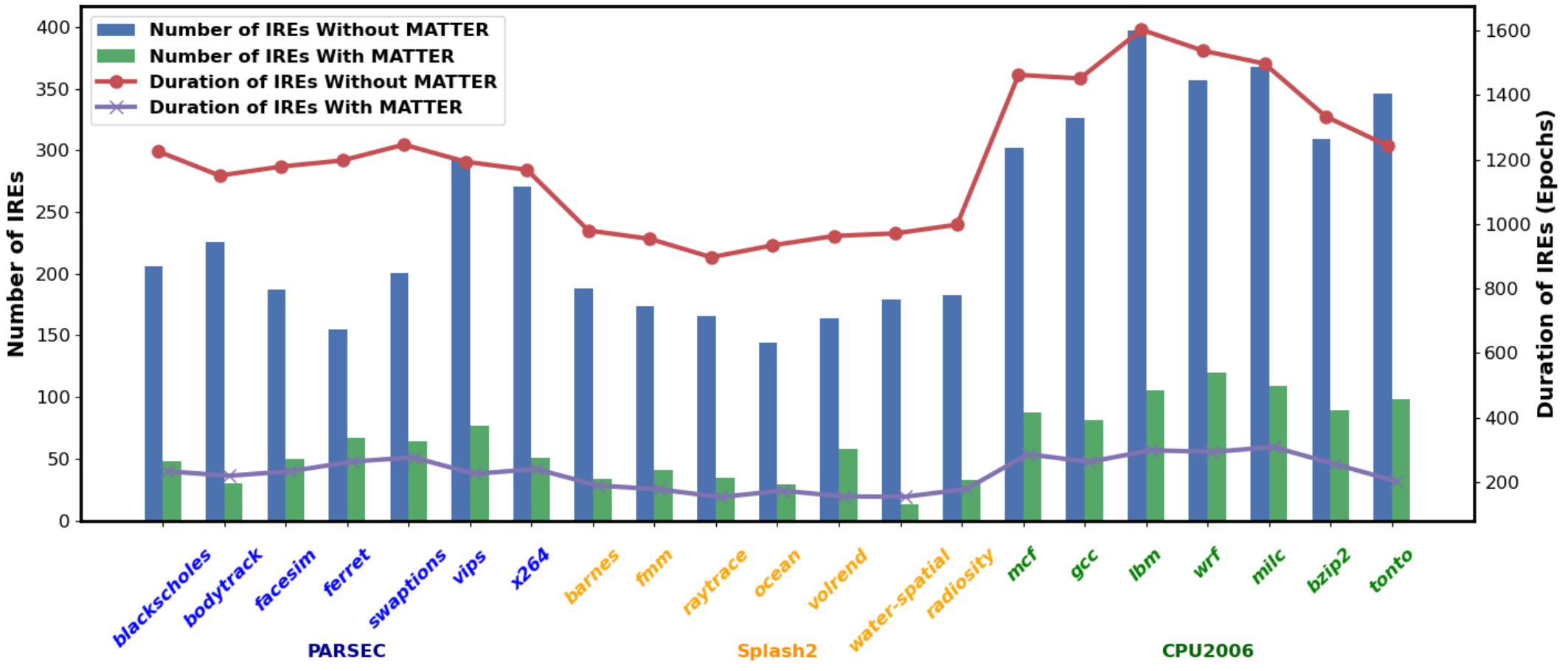} 
    \caption{Number and duration of IREs with and without MATTER across different benchmarks:  PARSEC (with simmedium input size), SPLASH2 (with small input size), and SPEC CPU2006
(with millions    of instructions) } 
    \label{fig:Throttle-res}
\end{figure*}


To evaluate the proposed attack and its impact on Multi-Processor System-on-Chips (MPSoCs) , we employ the open-source CoMeT thermal simulation toolchain \cite{kedia2022comet}. CoMeT provides a comprehensive framework for simulating performance, power, and thermal characteristics of multi/many-core processors and their memory systems. By integrating tools like Sniper\cite{carlson2011sniper} for performance simulation, McPAT \cite{li2009mcpat} for core power modeling, CACTI \cite{balasubramonian2017cacti} for memory power modeling, and HotSpot \cite{huang2006hotspot} for thermal simulation, CoMeT enables detailed interval simulations to analyze thermal interactions and evaluate thermal management strategies for various core-memory configurations. The simulated setup consists of a $45$ $nm$ Intel x86 Gainestown processor architecture \cite{intel}, running $4$ cores, with a frequency interval of $1.0$ $GHz$ - $4.0$ $GHz$ and a $0.1$ $GHz$ granularity. Furthermore, the CPU has three levels of cache memory and uses a DDR memory architecture of $1$ $GB$. 

In the following, we will analyze the degree to which the MATTER attack impacts the system's stability, performance (i.e., power consumption and core utilization), and stealthiness:

To thoroughly evaluate the performance impact of "MATTER" attack on mobile SoCs, we focus primarily on the frequency and duration of restricted epochs induced by the attack. We define these as Impactful Restricted Epochs (IREs). An IRE is characterized by a significant reduction in the operating frequency of a CPU core, initiated as a defensive response by DTM mechanism using DVFS technique. By tracking the number and duration of IREs, we quantify the impact of MATTER on system performance. We have defined IREs as they are critical preventive measures deployed by DVFS systems to mitigate the risk of thermal runaway (will be later discussed \ref{POW-Temp dynamics}) in mobile SoCs. These events play a vital role in enabling the DVFS to adjust performance parameters proactively, ensuring that the system temperature remains within safe operating limits.

\textbf{Setup environment:} to assess the magnitude of IRE reduction caused by MATTER attacks, we ran a series of tests using the Stanford Parallel Applications for Shared Memory 2 (SPLASH-2), Repository for Shared-Memory Computers (PARSEC) and CPU2006 benchmarks \cite{woo1995splash,bienia2008parsec}. The study involved seven distinct workloads for each benchmark including compute-intensive workloads (i.e., x264,GCC) and memory-intensive workloads (i.e., lbm,mcf) to assess the impact of workload scenarios on the the number of IREs.

\textbf{Experimental results.1:} Figure \ref{fig:Throttle-res} demonstrates that across a range of workloads, MATTER significantly disrupts DTM in mobile SoCs. On average, these attacks reduce DTM's ability to manage and balance thermal output effectively, resulting in an approximate 73\% degradation in DTM's optimal performance. This considerable drop in efficiency implies that MATTER attack framework disrupts the fundamental thermal balance strategies that DTM employs to protect the SoC from thermal overloads and ensure performance stability.

Specifically, DTM functions to throttle performance and initiate cooling mechanisms in response to rising temperatures. However, under the influence of MATTER's attack, these adaptive measures are compromised due to less IREs occured. MATTER leads DTM to either under-react. This miscalibration increases the likelihood of the SoC operating in sub-optimal conditions failing to engage cooling mechanisms when needed.

\begin{figure*}[t]
    \centering
    \includegraphics[scale=0.40]{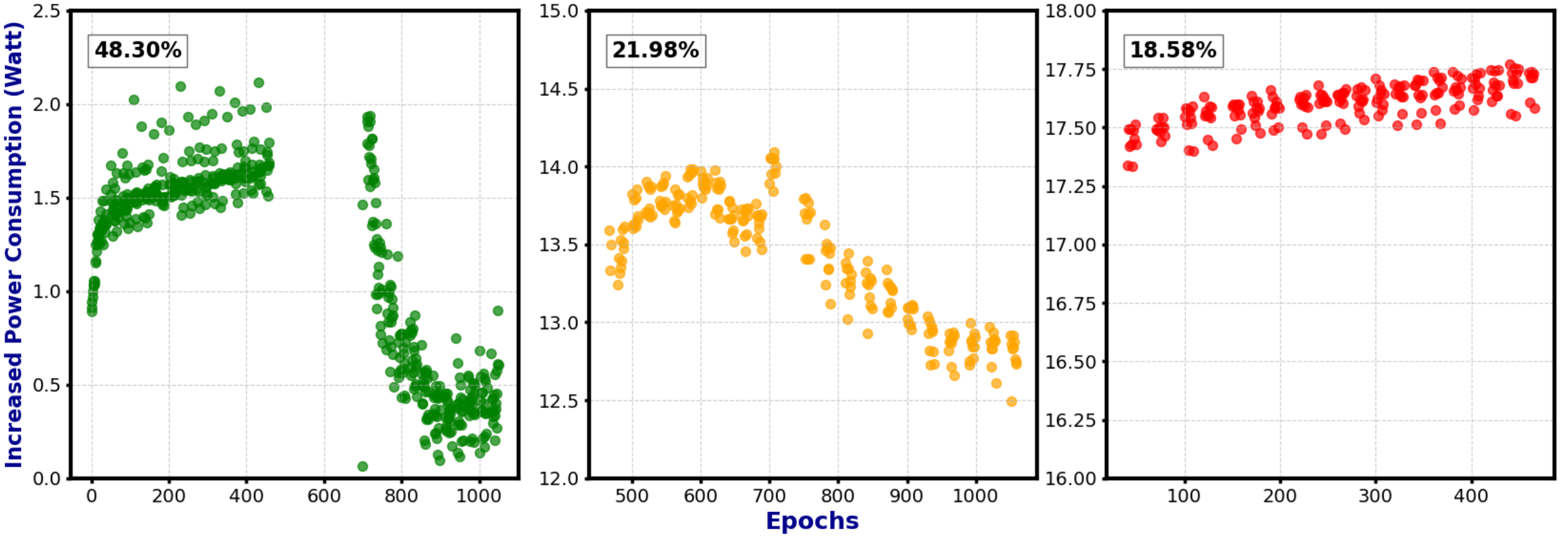}
    \caption{Increased Power Consumption Pre- and Post-Application of the MATTER Attack} 
    \label{fig:PowerDifference}
\end{figure*}

\begin{figure*}[t]
    \centering
        \includegraphics[scale=0.42]{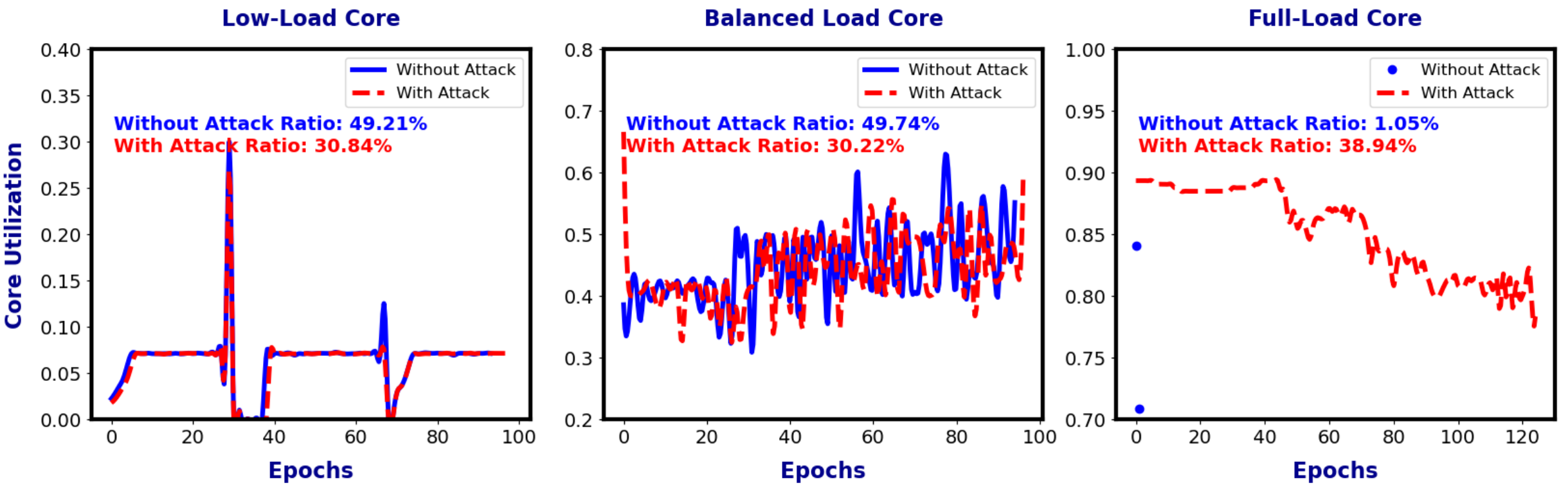}
    \vspace{-2mm}
    \caption{ Relative core resource utilization with and without exposure to the MATTER attack } 
    \label{fig:CoreUtilization}
    \vspace{-3mm}
\end{figure*}

\textbf{Experimental results.2:} The 73\% reduction in DTM efficiency demonstrates that the MATTER attack not only stresses the system but also depletes critical thermal management resources, undermining the SoC's ability to maintain optimal performance. By constantly triggering temperature fluctuations or causing inappropriate DTM responses, MATTER destabilizes the SoC's thermal environment. This disruption results in increased power demand as the DTM system struggles to keep temperatures within safe limits, leading to repeated thermal cycling. Such cycling accelerates wear on temperature-sensitive components like transistors, interconnects, and capacitors, contributing to the "aging" of the device. This aging manifests as reduced processing speed, lower energy efficiency, and a shortened device lifespan \cite{dadvar2005potential}. \Cref{fig:PowerDifference} illustrates the extra power consumed with and without the MATTER attack across mentioned mixed workloads in the setup environment section (x264, GCC, lbm, and mcf). Under complete execution of such mixed workloads, MATTER leads to a significant power increase, with 18.58\% of the execution time consuming $"17"$ watts or more, 21.98\% within the range of $"12-14"$ watts above baseline, and 48.30\% within a $"0-2.5"$ watt increase.

\subsection{Core Utilization}
\label{Utilization}

Core utilization refers to the extent to which a core is actively engaged in executing tasks. It is measured as a percentage of the total available processing time that the core is busy. Prolonged periods of high core utilization can significantly impact the thermal stability, power consumption, and overall reliability of mobile SoCs. When core utilization remains elevated for extended durations, it intensifies heat generation, leading to sustained high temperatures that challenge the effectiveness of DTM systems by increasing the likelihood of thermal runaway \cite{dadvar2005potential}.

In this experiment, we have used a mixed heterogeneous workload - loads include a mix of lbm, x264, exchange, and mcf benchmarks- in order to see the system's behavior under three different intensities i.e., low-load, balanced-load, and full-load conditions. we have used the same configuration outlined in \cref{Performance impact}. \Cref{fig:CoreUtilization} presents core utilization comparing "Without Attack" (blue) and "With Attack" (red) scenarios. In the absence of the MATTER attack, the core primarily operates at Low-Load (below 0.3) or Balanced Load (between 0.3 and 0.7) levels, indicating that the system remains predominantly in a stable, normal working condition. In contrast, with the MATTER attack applied, the time spent in the Full-Load condition (above 0.7) increases significantly (around 38\%), suggesting that the attack forces the core into higher utilization states more frequently, which could elevate the stress on the system and reduce its overall efficiency.

\subsection{Power-Temperature Stability Analysis}
\label{POW-Temp dynamics}
All the performance impacts shown in the subsections (\ref{Utilization} and \ref{Performance impact}) contribute to a critical threat known as thermal runaway, where the system experiences unstable conditions, escalating temperature uncontrollably. A critical metric for evaluating stability in this context is the analysis of fixed points in power-temperature dynamics and the risk of the thermal runaway \cite{bhat2017power}. A fixed point refers to a stable equilibrium where power consumption and temperature maintain a steady state, allowing the system to operate without further fluctuations. At this equilibrium, minor perturbations do not cause deviations, ensuring the system's resilience against instability. However, the MATTER attack can significantly impact system's equilibrium and create intentional thermal runaways.

\begin{figure}[t]
    \centering
    \includegraphics[width=0.50\textwidth]{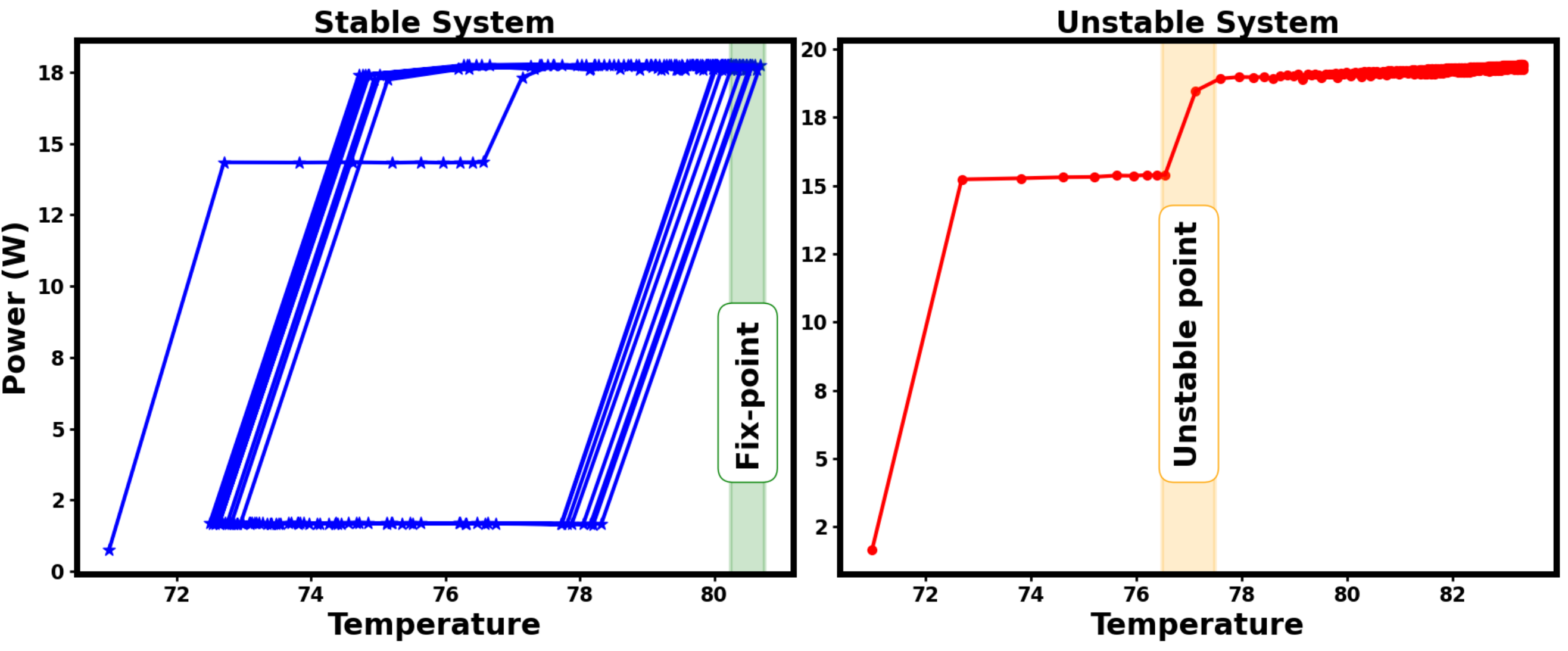} 
    \vspace{-4mm} 
    \caption{Power-Temperature Dynamics of the stable/unstable system} 
    \label{fig:stabilityAnalysis}
    \vspace{-6mm} 
\end{figure}

For stability analysis, we execute the $FFT$ benchmark from the SPLASH-2 suite, chosen for its ability to simulate representative workloads and generate elevated temperature levels. This benchmark effectively stresses the thermal behavior of the system, allowing us to observe temperature dynamics under high load. 
This setup facilitates observing how the target core (core $"1"$ in this case) responds under thermal pressure, particularly in relation to DTM policies. The simulation is configured as outlined in \cref{Performance impact}. During this simulation, DTM is activated to modulate thermal responses in real time, providing insights into how effectively it manages temperatures under intensive conditions and the MATTER attack. \Cref{fig:stabilityAnalysis} presents the simulation results of the system under normal and attacked operating conditions. The stable graph (blue) is shown to understand baseline stability by demonstrating a stable system where a single, well-defined fix-point is achieved. At this point, the power curve-which represents the system's heat generation-intersects the temperature curve at a single equilibrium point. This intersection represents the system's natural balance, where the generated heat is effectively dissipated, allowing the temperature to remain stable.

A key observation is that the slope of the power curve at the equilibrium point is less steep than the slope of the temperature curve. This slope relationship is significant as any minor deviations from this fix-point trigger corrective forces, effectively returning the system toward equilibrium. As such, even if small fluctuations in power or temperature occur, the system is inherently designed to stabilize, preventing any sustained deviation from its steady-state condition. This stability profile, established without any thermal attack, confirms the robustness of the system's DTM mechanisms under regular conditions. It illustrates that the DTM can manage typical variations in workload and heat dissipation, maintaining consistent performance without triggering extreme temperature variations or instability. 

In contrast,  the right panel in\Cref{fig:stabilityAnalysis} illustrates the post-attack behavior of the system, which now exhibits characteristics of thermal instability and potential thermal runaway. After the attack is introduced, there is an area (interval)- shown by the orange color- that represents a critical instability threshold. When the system operates above these temperatures, it is highly susceptible to thermal runaway, as the attack amplifies the tendency for temperature to increase uncontrollably. Under these conditions, any deviation that raises the temperature above this threshold leads the system toward the upper point, where temperature escalation accelerates. 
The system's high sensitivity to even minor changes in power or temperature in this post-attack state makes it especially vulnerable. 
This heightened sensitivity suggests that the DTM mechanisms, effective under normal conditions, struggle to counteract the thermal stresses under the introduced attack.

\begin{table}[t]
\small 
\setlength{\tabcolsep}{1pt} 
\caption{Average Temperature of Trojan-Free and Trojan-infected IC under different configurations}
\label{tab:AverageFrequency}
\begin{tabular}{|c|c|c|c|}
\hline
\rowcolor[HTML]{EFEFEF} 
\textbf{Configuration} & Trojan-Free & Trojan-Infected & Deviation \\ \hline
gainestown\_DDR-4core & 77.54\degree C & 78.07\degree C & -0.6\% \\ \hline
\rowcolor[HTML]{EFEFEF} 
gainestown\_DDR\_16core & 78.36\degree C & 78.93\degree C & -0.7\% \\ \hline
gainestown\_16channel\_32cores & 78.74\degree C & 77.74\degree C & +1.2 \\ \hline
\rowcolor[HTML]{EFEFEF} 
gainestown\_16channel\_48cores & 79.02\degree C & 78.84\degree C & -0.22\% \\ \hline
\end{tabular}
\end{table}

\subsection{Stealthiness} 
The evaluation of MATTER's stealthiness centers on its detection rate, offering a robust measure of the attack's inherent weaknesses in stealthiness. Given that MATTER manipulates temperature responses, a straightforward detection approach based on the average temperature of Trojan-free ICs is effective in identifying infected ICs. 
To systematically evaluate the stealthiness of thermal attacks, we have employed four distinct system configurations, varying in core count from 4 to 48 cores. These configurations, as outlined in \Cref{tab:AverageFrequency}, facilitate a thorough examination of temperature responses across systems of different scales. By testing a range of configurations, we aim to capture the nuanced impact of thermal attacks on system behavior, taking into account how the number of cores influences thermal dynamics, performance, and potential detection thresholds. 

Findings presented in \Cref{tab:AverageFrequency}  demonstrate the resilience of MATTER against detection methods based on average temperature analysis. The stealthiness achieved in all simulated configurations underscores the effectiveness of the attack in bypassing basic detection techniques, which rely on identifying deviations in average temperature that would indicate anomalous behavior.

%% file: Sources/5-ConC.tex
\section{Conclusions}
\label{Sec:ConC}
This paper uncovered a thermal vulnerability in mobile System-on-Chips that rely on dynamic thermal management for their power/thermal fluctuations. The vulnerability was used to define a new attack vector that targets device's thermal management unit to degrade performance by up to 73\%. By sending fake temperature signals, the proposed thermal attack can cause a device to make wrong decisions about managing its performance, which could also alter the system's reliability over time. Unlike previous attacks that target temperature sensors directly, the proposed attack does not touch the sensor circuitry as it sits at the  interface between temperature sensor and the device's thermal management unit. We anticipate that this make the attack harder to detect as it can be defined either as a hardware Trojan at the hardware level or as a software Trojan at the operating system's kernel. Our findings reveal a significant weakness in how mobile devices protect themselves from such thermal attacks, suggesting that current technology is not prepared for these sophisticated threats.  